\documentclass[aps, prd
, twocolumn
, nofootinbib
, superscriptaddress
]{revtex4-2}

\usepackage{graphicx}
\usepackage{xcolor}
\usepackage{mathrsfs,mathtools}
\usepackage{physics,amssymb}
\usepackage{siunitx}
\usepackage{bm}
\usepackage{braket}
\usepackage{cases}
\usepackage{comment}
\usepackage{soul}
\usepackage{cancel}
\usepackage{url}
\usepackage{xspace}
\usepackage{acronym}
\usepackage{siunitx}
\usepackage[normalem]{ulem}

\usepackage[colorlinks=true
,urlcolor=DARKBLUE
,anchorcolor=DARKBLUE
,citecolor=DARKBLUE
,filecolor=DARKBLUE
,linkcolor=DARKBLUE
,menucolor=DARKBLUE
,linktocpage=true
,pdfproducer=medialab
,pdfa=true
]{hyperref}

\DeclareMathOperator{\erfc}{erfc}

\newcommand{\ee}{\mathrm{e}}

\newcommand{\cs}{c_{\mathrm{s}}}
\newcommand{\csmin}{c_{\mathrm{s},\mathrm{min}}}
\newcommand{\GW}{\mathrm{GW}}
\newcommand{\CMB}{\mathrm{CMB}}
\newcommand{\obs}{\mathrm{obs}}

\newcommand{\uc}{\mathrm{c}}

\newcommand{\uf}{\mathrm{f}}

\newcommand{\calH}{\mathcal{H}}

\newcommand{\calN}{\mathcal{N}}

\newcommand{\calP}{\mathcal{P}}
\newcommand{\calR}{\mathcal{R}}

\newcommand{\bfk}{\mathbf{k}}

\newcommand{\calT}{\mathcal{T}}

\newcommand{\ur}{\mathrm{r}}

\newcommand{\us}{\mathrm{s}}

\newcommand{\beae}[1]{\begin{equation}\begin{aligned} #1 \end{aligned}\end{equation}}
\newcommand{\bege}[1]{\begin{equation}\begin{gathered} #1 \end{gathered}\end{equation}}
\newcommand{\bae}[1]{\begin{align} #1 \end{align}}

\newcommand{\bfe}[4]{
\begin{figure} 
	\centering
	\includegraphics[#1]{#2}
	\caption{#3}
	\label{#4}
\end{figure}}
\newcommand{\bme}[1]{\begin{multline} #1 \end{multline}}

\definecolor{MONZA}{HTML}{CF000F}
\definecolor{DARKBLUE}{HTML}{00008b}
\definecolor{DARKMAGENTA}{HTML}{8b008b}
\definecolor{DARKRED}{HTML}{8b0000}

\begin{document}
\title{The LISA forecast on a smooth crossover beyond the Standard Model through the scalar-induced gravitational waves}
\date{\today}

\author{Albert Escriv\`a}
\email{escriva.manas.albert.y0@a.mail.nagoya-u.ac.jp}
\affiliation{Division of Particle and Astrophysical Science, Graduate School of Science, 
Nagoya University, Nagoya 464-8602, Japan}
\author{Ryoto Inui}
\email{inui.ryoto.a3@s.mail.nagoya-u.ac.jp}
\affiliation{Division of Particle and Astrophysical Science, Graduate School of Science, 
Nagoya University, Nagoya 464-8602, Japan}
\author{Yuichiro Tada}
\email{tada.yuichiro.y8@f.mail.nagoya-u.ac.jp}
\affiliation{Institute for Advanced Research, Nagoya University,
Furo-cho Chikusa-ku, 
Nagoya 464-8601, Japan}
\affiliation{Division of Particle and Astrophysical Science, Graduate School of Science, 
Nagoya University, Nagoya 464-8602, Japan}
\author{Chul-Moon Yoo}
\email{yoo.chulmoon.k6@f.mail.nagoya-u.ac.jp}
\affiliation{Division of Particle and Astrophysical Science, Graduate School of Science, 
Nagoya University, Nagoya 464-8602, Japan}

\begin{abstract}
Supposing the \ac{LISA} \ac{GW} detector, we exhibit the detectability of a hypothetical smooth crossover in the early universe beyond the Standard Model of particle physics through the \ac{SIGW} in terms of the Fisher forecast. A crossover at $\sim\SI{100}{TeV}$ can leave a signal on the \ac{GW} spectrum in the $\sim\si{mHz}$ frequency range, the sweet spot of the \ac{LISA} sensitivity.
These possibilities are also interesting in the \ac{PBH} context as the associated \ac{PBH} mass $\sim10^{22}\,\si{g}$ lies at the window to explain the whole dark matter.
We found that the properties of the crossover can be well determined if the power spectrum of primordial scalar perturbations are as large as $\sim\num{5e-4}$ 
on the corresponding scale $\sim10^{12}\,\si{Mpc^{-1}}$.
\end{abstract}

\maketitle

\acrodef{PBH}{primordial black hole}
\acrodef{CMB}{cosmic microwave background}
\acrodef{LISA}{Laser Interferometer Space Antenna}
\acrodef{DECIGO}{DECi-hertz Interferometer Gravitational-wave Observatory}
\acrodef{PDF}{probability density function}
\acrodef{EoM}{equation of motion}
\acrodef{EoS}{equation-of-state}
\acrodef{GW}{gravitational wave}
\acrodef{RD}{radiation-dominated}
\acrodef{DM}{dark matter}
\acrodef{PTA}{pulsar timing array}
\acrodef{TDI}{Time Delay Interferometry measurement }
\acrodef{IMS}{Interferometry Metrology System}
\acrodef{SGWB}{stochastic gravitational wave background}
\acrodef{SIGW}{scalar-induced gravitational wave}
\acrodef{SNR}{signal to noise ratio}
\acrodef{SM}{Standard Model}
\acrodef{SC}{smooth cross-over}
\acrodef{TDI}{Time Delay Interferometry}
\acrodef{QCD}{quantum chromodynamics}

\acresetall
\section{Introduction}

The Standard Model of particle physics is one of the greatest successes in the history of physics.
It unifies all known elementary particles in a mathematically simple way and exhibits high predictability in observables with extraordinary precision represented by the electron's anomalous magnetic dipole moment~\cite{Kusch:1948mvb,Schwinger:1948iu}.
It is, however, true that some phenomena beyond the Standard Model such as the non-zero neutrino mass~\cite{Super-Kamiokande:1998kpq}, the baryon asymmetric universe~\cite{Sakharov:1967dj}, dark matter~\cite{1933AcHPh...6..110Z}, etc. have been clarified and hence the Standard Model must be improved in these respects.
Each hierarchy of physics is often divided by phase transitions or crossovers, as seen in the Standard Model.
As in the case of crossovers in the electroweak~\cite{Kajantie:1996mn,Laine:1998vn,Rummukainen:1998as} and \ac{QCD}~\cite{Aoki:2006we}, it may be natural to expect a novel smooth crossover beyond the current energy frontier, e.g., a few $\si{TeV}$ or higher.\footnote{See, e.g., Ref.~\cite{Lu:2022yuc} for a first-order extension of \ac{QCD} in the context of \acp*{PBH} and \acp*{GW} and Ref.~\cite{Athron:2023xlk} for a review on \acs*{GW} production in a first-order phase transition.}

On the other astrophysical side, the first direct detection of \acp{GW} by the LIGO collaboration~\cite{LIGOScientific:2016aoc} opened a new era to explore physics. 
The LIGO--Virgo--KAGRA collaborations detected $\sim90$ \ac{GW} events from compact binary coalescences up until the third observing run~\cite{KAGRA:2021vkt}. 
In addition, the international \ac{PTA} networks recently reported the evidence of the \acp{SGWB}~\cite{NANOGrav:2023gor, EPTA:2023fyk, Reardon:2023gzh, Xu:2023wog}. 
Although their standard interpretation is the superposition of \acp{GW} from supermassive black hole binary systems~\cite{Antoniadis:2022pcn, Reardon:2023gzh, Xu:2023wog}, they or a part of them can originate from the early universe physics (see, e.g., Refs.~\cite{NANOGrav:2023hvm, EPTA:2023xxk}), represented by the preheating~\cite{Starobinsky:1979ty,Turner:1996ck, Khlebnikov:1997di,Garcia-Bellido:2007fiu,Aggarwal:2020olq}, cosmic strings~\cite{LIGOScientific:2017ikf,LIGOScientific:2021nrg,vanRemortel:2022fkb}, bubble collisions at a first-order phase transition~\cite{Witten:1984rs,Kosowsky:1991ua,Kosowsky:1992vn,Caprini:2018mtu}, and so on. 

The survey of the effective degrees of freedom of particles in the early universe \emph{through} \acp{SGWB} has also been investigated (see, e.g., Refs.~\cite{Kuroyanagi:2008ye,Saikawa:2018rcs} for the detailed studies on the impact of the Standard Model physics on the spectrum of the primordial \ac{GW} background, Ref.~\cite{Franciolini:2023wjm} for the \ac{QCD} effect on the so-called \emph{causality tail} $\propto f^{3+2\frac{3w-1}{3w+1}}$ of the superHubble \ac{GW}s and its implication on the $\si{nHz}$ \ac{GW} background reported by the \ac{PTA} collaborations, etc.).
Cosmic inflation is expected to have generated almost scale-invariant \acp{SGWB} from the quantum vacuum fluctuations of the tensor modes but their relic spectrum depends on the thermal history of the universe.
If some particles get massive and become non-relativistic in a hot universe, their pressure reduces and hence the dilution rate of their energy density decreases.
Accordingly, the relative energy density of \acp{SGWB} to the background energy density decreases. One sees that the equation of motion for \acp{GW}~\eqref{eq: ptb EoM} indeed depends on the \ac{EoS} parameter $w=p/\rho$ of the background energy density $\rho$ and pressure $p$.
The particle dropout from the thermal plasma is recorded as a \emph{break} of the \ac{GW} spectrum.

\bfe{width=0.9\hsize}{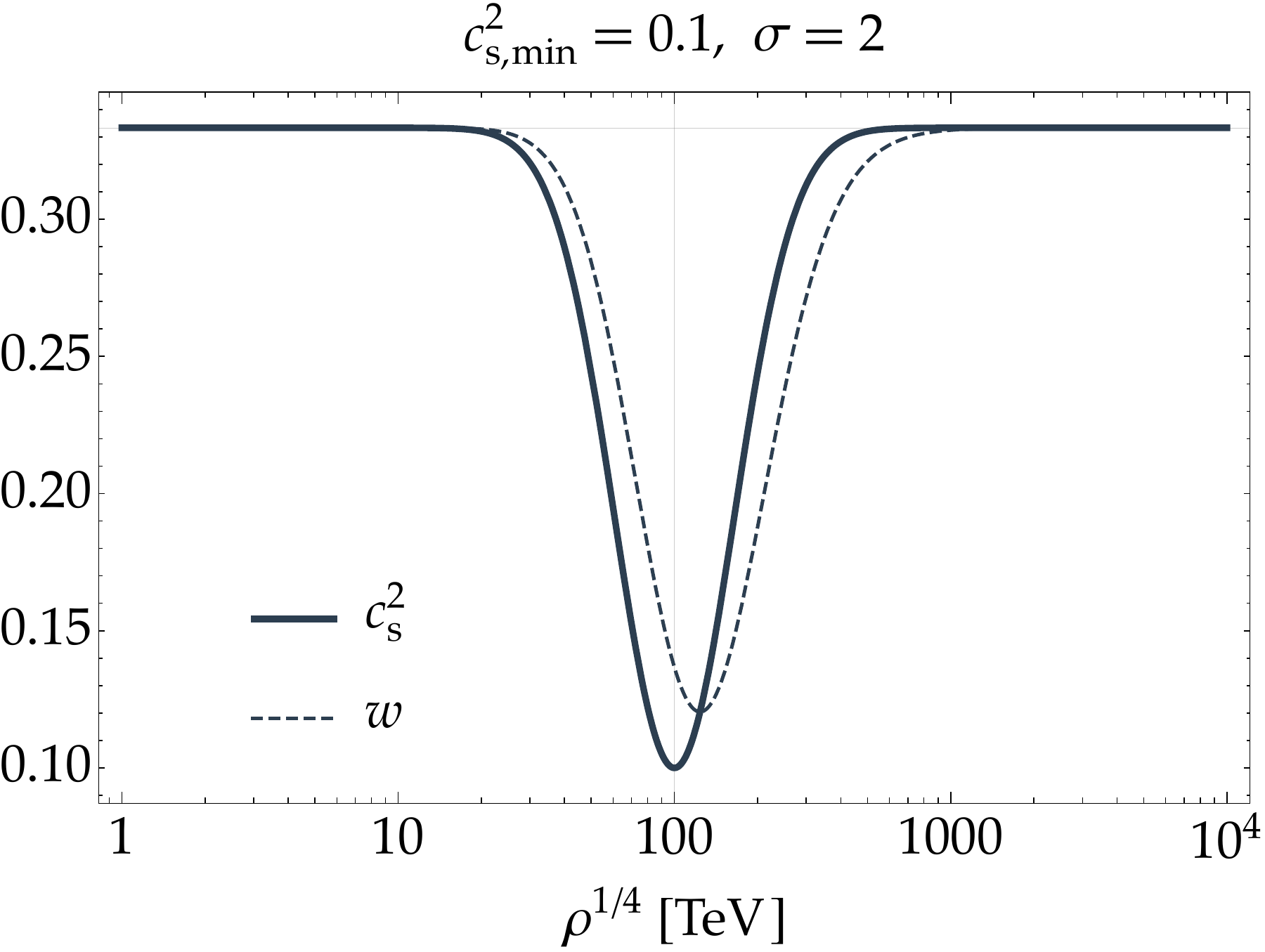}{Illustrations of the modelings of the sound speed squared $\cs^2$~\eqref{eq: cs2} (black thick) and the \ac{EoS} parameter $w$~\eqref{eq: w} (dashed) for $\rho_*^{1/4}=\SI{100}{TeV}$ (vertical thin), $\csmin^2=0.1$, and $\sigma=2$. The horizontal thin line indicates the exact radiation value $w=\cs^2=1/3$.}{fig: cs2 and w}

\begin{figure*}
    \centering
    \begin{tabular}{c}
        \begin{minipage}[b]{0.5\hsize}
            \centering
            \includegraphics[width=0.9\hsize]{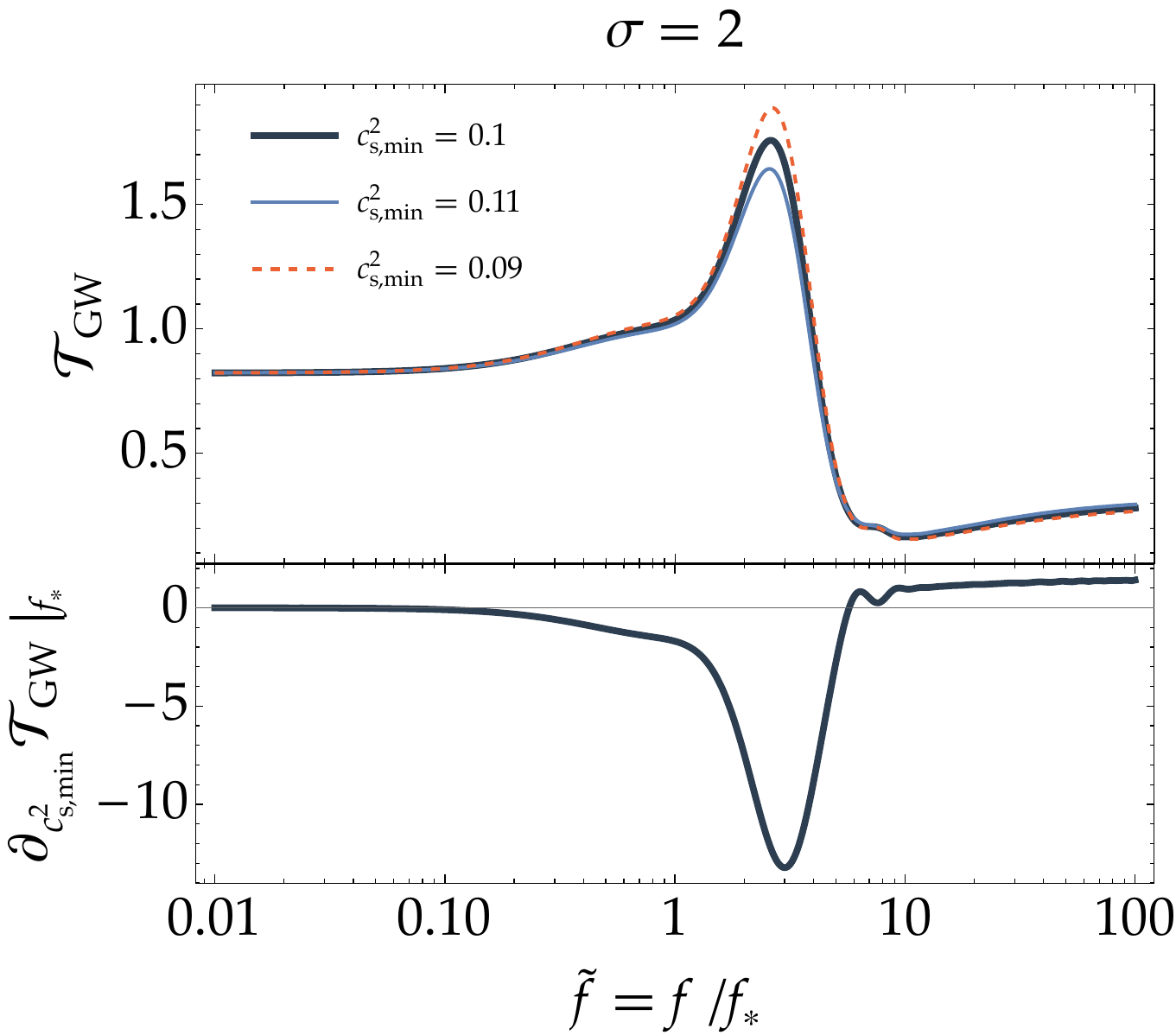}
        \end{minipage}
        \begin{minipage}[b]{0.5\hsize}
            \centering
            \includegraphics[width=0.9\hsize]{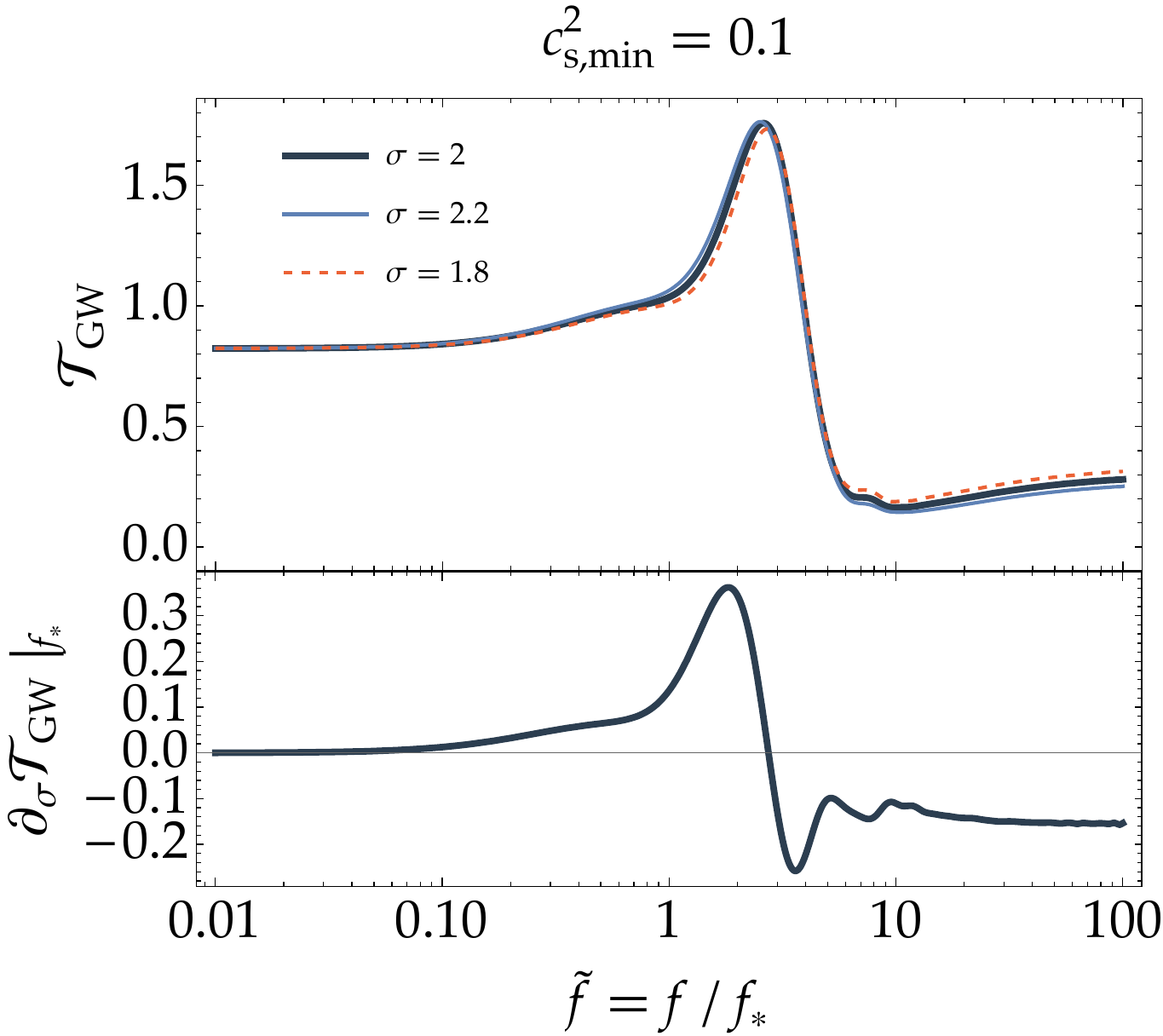}
        \end{minipage}
    \end{tabular}
    \caption{The numerically-calculated \ac{GW} template $\calT_\GW(\tilde{f})$~\eqref{eq: GW_model} with varying $\csmin^2$ (left) and $\sigma$ (right).
    The bottom panels show its derivatives in these parameters obtained by the finite difference approximation~\eqref{eq: finite difference}.}
    \label{fig: template}
\end{figure*}

Recently, another type of \acp{SGWB}, called the \acp{SIGW}, has attracted attention.
The primordial curvature perturbations can source \acp{SGWB} through the second-order interaction between the scalar and tensor metric perturbations and could be detected if the scalar curvature perturbations were significantly enhanced compared to the observed value on a cosmological scale (see, e.g., Refs.~\cite{Ananda:2006af, Baumann:2007zm, Saito:2008jc, Espinosa:2018eve, Kohri:2018awv, Domenech:2021ztg}).
As the scalar perturbation depends not only on the \ac{EoS} parameter $w$ but also the plasma sound speed $\cs^2=\pdv*{p}{\rho}$, \acp{SIGW} are sensitive both on them and helpful to probe the thermal history in more detail (see Refs.~\cite{Hajkarim:2019nbx,Domenech:2019quo,Domenech:2020kqm,Abe:2020sqb,Abe:2023yrw,Escriva:2023nzn}).
Supposing a crossover around $\sim100\,\si{TeV}$ beyond Standard Model, the characteristic frequency of the modulated \acp{SIGW} reads $\sim\si{mHz}$ which is an interesting target of future space-based gravitational wave detectors such as \ac{LISA}~\cite{LISA:2017pwj} and \ac{DECIGO}~\cite{Seto:2001qf, Kawamura:2011zz}, and furthermore large scalar perturbations can bring about the \acp{PBH}~\cite{Zeldovich:1967lct,Hawking:1971ei,Carr:1974nx,Carr:1975qj} (see also Refs.~\cite{Carr:2020gox,Escriva:2022duf,Yoo:2022mzl,Carr:2023tpt} for recent review articles) of $\sim10^{22}\,\si{g}$ as an intriguing candidate of dark matter in the presence of crossovers~\cite{Escriva:2022yaf,
Escriva:2023nzn}. 
In this \emph{Letter}, we want to explicitly show the detectability of that hypothetical smooth crossover beyond the Standard Model in the LISA frequency range. For that purpose, we phenomenologically parameterize the reduction of the sound speed from the relativistic value $\cs^2=1/3$ due to a hypothetical crossover as a function of the energy density $\rho$ of the background fluid by (see Fig.~\ref{fig: cs2 and w})
\bae{\label{eq: cs2}
    \cs^2(\rho)=\frac{1}{3}-\pqty{\frac{1}{3}-\csmin^2}\exp[-\frac{\ln[2](\rho/\rho_*)}{2\sigma^2}],
}
with three model parameters $\bm{\theta}=(\ln\rho_*^{1/4},\csmin^2,\sigma)$ as in Ref.~\cite{Escriva:2023nzn}, and perform the Fisher forecast on these parameters via \acp{SIGW} with the one-year \ac{LISA} observation, assuming a scale-invariantly amplified scaler power spectrum
\bae{\label{eq: calPzeta}
    \calP_\zeta(k)=\frac{k^3}{2\pi^2}\abs{\zeta_k}^2=A_\us\gg\calP_\zeta(k_\CMB)\sim\num{2e-9},
}
for the primordial curvature perturbation $\zeta$, for which the change in the spectrum of the induced GWs can be only associated with the effect caused by the crossover.
Throughout the \emph{Letter}, we adopt the Planck unit $c=\hbar=G=1$.

\section{Scalar-induced gravitational waves through a smooth crossover}

Supposing the functional form of $\cs^2$ as Eq.~\eqref{eq: cs2}, the consistent \ac{EoS} parameter is solved as
\bae{\label{eq: w}
    &w(\rho)=\frac{1}{\rho}\int_0^{\tilde{\rho}}\cs^2(\tilde{\rho})\dd{\tilde{\rho}} \nonumber \\
    &=\frac{1}{3}-\frac{\sigma}{\rho/\rho_*}\sqrt{\frac{\pi}{2}}\ee^{\sigma^2/2}\pqty{\frac{1}{3}-\csmin^2}\erfc\qty[\frac{\sigma^2-\ln\rho/\rho_*}{\sqrt{2}\sigma}],
}
with the complementary error function $\erfc(z)=\frac{2}{\sqrt{\pi}}\int_z^\infty\ee^{-t^2}\dd{t}$.
The background fluid dynamics is fixed by solving the evolution equation for $\rho$ and the Hamiltonian constraint for the scale factor $a$:
\bae{\label{eq: bg EoM}
    \partial_\eta\rho=-\sqrt{24\pi}\qty(1+w(\rho))a\rho^{3/2} \qc 
    \partial_\eta a=\sqrt{\frac{8\pi\rho}{3}}a^2,
}
with the conformal time $\eta=\int a^{-1}\dd{t}$.
Hereafter, we also use the conformal Hubble parameter defined by $\calH=a'/a$ where the prime denotes the derivative in the conformal time.

\begin{figure}
    \centering
    \includegraphics[width=\hsize]{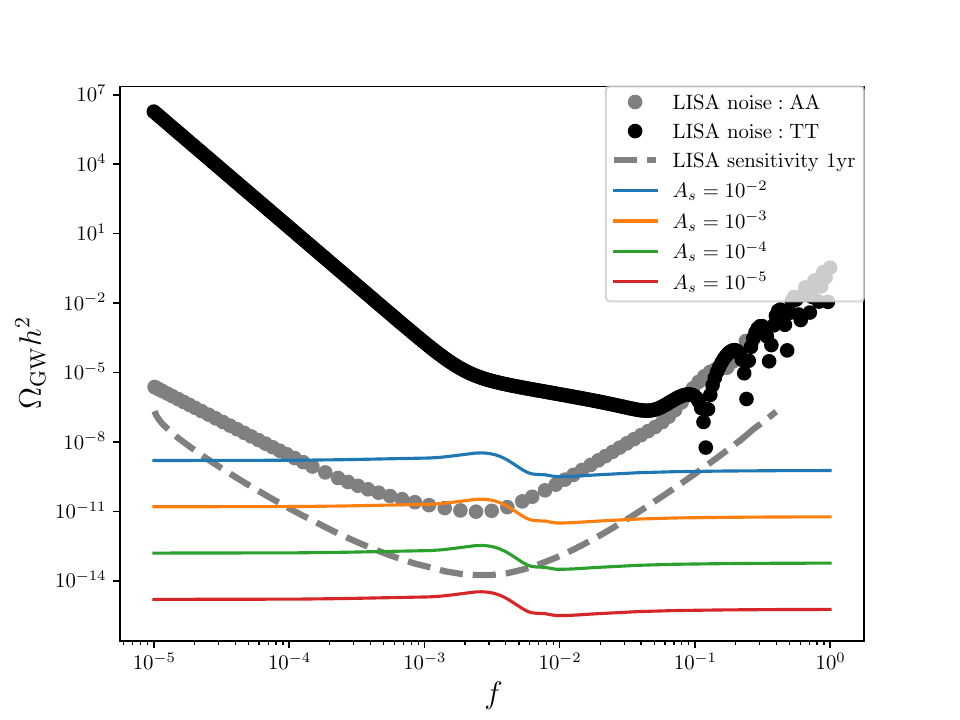}
    \caption{The \ac{GW} energy density parameter $\Omega_\GW h^2$ and the LISA noise energy density parameters $\Omega_{n,AA}h^2=\Omega_{n,EE}h^2$ (gray dots) and $\Omega_{n,TT}h^2$ (black dots). 
    The gray dashed line, as a reference, denotes the LISA's sensitivity for the power-law \ac{SIGW} assuming one-year observations~\cite{Schmitz:2020syl}. The color lines represent the \ac{SIGW} spectrum for each $A_\us$.}
    \label{fig: Omg_vs_LISA}
\end{figure}

Over the background evolution, one finds the dynamics of the scalar and tensor perturbations via the equations of motion (see, e.g., Ref.~\cite{Mukhanov:2005sc}), 
\bege{\label{eq: ptb EoM}
    \Phi_k''+3\calH(1-\cs^2)\Phi_k'+\bqty{\cs^2k^2+3\calH^2(\cs^2-w)}\Phi_k'=0, \\
    \pqty{\partial_\eta^2+k^2-\frac{1-3w}{2}\calH^2}g_{ik}=0,
}
for the scalar transfer function $\Phi_k$ and two independent tensor mode functions $g_{1k}$ and $g_{2k}$. The scalar transfer function should be initialized by $\Phi_k(\eta\to0)\sim1$, while one can take two distinct but arbitrary initial conditions for $g_{1k}$ and $g_{2k}$ such as $g_{1k}(\eta\to0)\sim1$ and $g_{2k}(\eta\to0)\sim k\eta$.
The power spectrum of \acp{SIGW} is calculated in the Green's function method as 
\bme{
    \calP_h(k,\eta)=\frac{64}{81a^2(\eta)}\int_{\abs{\bfk_1-\bfk_2}\leq k\leq\abs{\bfk_1+\bfk_2}}\hspace{-50pt}\dd{\ln k_1}\dd{\ln k_2}I^2(\eta,k_1,k_2,k) \\
    \times\frac{\left(k_1^2  - (k^2 - k_2^2 + k_1^2 )^2/(4k^2)\right)^2}{k_1k_2k^2} \calP_{\zeta}(k_1)\calP_{\zeta}(k_2),
}
with the kernel
\bae{
    I(\eta,k_1,k_2,k)=k^2\int^\eta\dd{\tilde{\eta}}G_\bfk(\eta,\tilde{\eta})a(\tilde{\eta})f(k_1,k_2,\tilde{\eta}),
}
where the tensor Green's function is given by
\bae{
    G_\bfk(\eta,\tilde{\eta})=\frac{1}{\calN_k}\bqty{g_{1k}(\eta)g_{2k}(\tilde{\eta})-g_{1k}(\tilde{\eta})g_{2k}(\eta)}\Theta(\eta-\tilde{\eta}),
}
with the normalization\footnote{It is indeed time-independent as ensured by the equation of motion~\eqref{eq: ptb EoM}.} $\calN_k=g_{1k}'(\tilde{\eta})g_{2k}(\tilde{\eta})-g_{1k}(\tilde{\eta})g_{2k}'(\tilde{\eta})$
and the source function reads
\bme{
    f(k_1,k_2,\eta) = 2\Phi_{k_1}(\eta)\Phi_{k_2}(\eta) \\ 
    + \frac{4}{3(1\!-\!w(\eta))} \!\left(\!\Phi_{k_1}\!(\eta)+ \frac{\Phi^{\prime}_{k_1}\!(\eta)}{\calH(\eta)}\!\right)\!\left(\!\Phi_{k_2}\!(\eta)+ \frac{\Phi^{\prime}_{k_2}\!(\eta)}{\calH(\eta)}\!\right).
}
The current \ac{GW} energy density parameter $\Omega_{\GW}$ is given by 
\bae{
    \label{eq: current_GW_energy}
    \Omega_{\GW}(k, \eta_0)h^2 = \Omega_{\rm r, 0}h^2\frac{1}{24}\left(\frac{a_{\rm c} \calH_{\rm c}}{a_{\rm f} \calH_{\rm f}}\right)^2 \left(\frac{k}{\calH_{\rm c}}\right)^2\overline{\calP_{h}(k, \eta_{\rm c})},
}
where $\Omega_{\ur,0}$ is the current radiation energy density parameter, $\Omega_{\ur, 0}h^2 = 4.2 \times 10^{-5}$, with the dimensionless Hubble parameter $h = H_0/(100\,\si{km.s^{-1}.Mpc^{-1}})$, the subscript ``$\uc$" denotes the time well after the horizon reentry when the \ac{GW} density parameter becomes almost constant, and the subscript ``$\uf$" indicates the time when the degrees of freedom of relativistic species and the entropy are well reduced to the current values after all phase transitions but still in the radiation-dominated era. $\overline{\calP_{h}(k, \eta_{\uc})}$ is the time average of $\calP_h(k)$ around $\eta_\uc$ within a period of the oscillation. We use the \textit{IGWsfSC} code~\cite{IGWsfSCcc} to numerically compute $\Omega_{\GW}(k, \eta_0)$ through the modeled crossover~\eqref{eq: cs2} and \eqref{eq: w}.

It is useful to make a template as
\bae{
    \label{eq: GW_model}
    \Omega_\GW(f)h^2=\Omega_{\ur0}h^2A^2_\us \calT_{\GW}(\tilde{f}),
}
where the template $\calT_\GW(\tilde{f})$ should be a function only of the renormalized frequency $\tilde{f}=f/f_*$ with the characteristic frequency $f_*=k_*/(2\pi)=\calH_*/(2\pi)$, which corresponds to the Hubble scale at the time when $\rho=\rho_*$.
$f_*$ weakly depends also on $\csmin^2$ and $\sigma$ through the background equation~\eqref{eq: bg EoM}.
Supposing $a=\num{1.93e-17}$ and $\eta=\SI{1.40e-11}{Mpc}$ at $\rho^{1/4}=10\,\si{TeV}$ to be consistent with the thermal history in the Standard Model and the normalization $a=1$ today (we have utilized the fitting formulae for effective degrees of freedom given in Ref.~\cite{Saikawa:2018rcs}), one can numerically find the relation (see Appendix~\ref{sec: rho and f}),
\bae{\label{eq: f* fitting}
    \frac{f_*}{\si{mHz}}=\bqty{c_1+c_2(\csmin^2-0.1)+c_3(\sigma-2)}\frac{\rho_*^{1/4}}{\SI{100}{TeV}},
}
with
\bae{
    c_1=1.019 \qc
    c_2=0.3898 \qc
    c_3=-0.0579,
}
around the fiducial values $\csmin^2=0.1$ and $\sigma=2$.
The partial derivative of the \ac{GW} spectrum in $\rho_*^{1/4}$ fixing $\csmin^2$ and $\sigma$, which is necessary for the Fisher analysis, is equivalent to that in $f_*$ as
\bae{
    \eval{\partial_{\ln\rho_*^{1/4}}\Omega_\GW(f)h^2}_{\csmin^2,\sigma}=-\Omega_{\ur0}h^2A_\us^2\tilde{f}\calT_\GW'(\tilde{f}),
}
where $\calT_\GW'(\tilde{f})=\partial_{\tilde{f}}\calT_\GW(\tilde{f})$.

The examples of templates for $\csmin^2=0.1\pm0.01$ and $\sigma=2\pm0.2$ are shown in Fig.~\ref{fig: template}. 
Such templates with slightly shifted parameters around the fiducial values enable us to evaluate other derivatives:
\bae{\label{eq: finite difference}
    \eval{\partial_{\theta_i}\calT_\GW(\tilde{f})}_{f_*}\!\!\!\approx\frac{\calT_\GW(\tilde{f}\mid\theta_i+\Delta\theta_i)-\calT_\GW(\tilde{f}\mid\theta_i-\Delta\theta_i)}{2\Delta\theta_i},
}
for $\theta_2=\csmin^2$ and $\theta_3=\sigma$.
Note that, due to the weak dependence of $f_*$ on $\csmin^2$ and $\sigma$, the derivatives fixing $\rho_*$ is contaminated by the derivative in $f_*$ as
\bae{
    \eval{\partial_{\theta_i}\calT_\GW(\tilde{f})}_{\rho_*}=\eval{\partial_{\theta_i}\calT_\GW(\tilde{f})}_{f_*}-\frac{c_i}{c_1}\tilde{f}\calT_\GW'(\tilde{f}),
}
for $i=2$ and $3$.

\section{Fisher analysis in the LISA system}

\begin{figure*}
    \centering
    \begin{tabular}{c}
        \begin{minipage}{0.5\hsize}
            \centering
            \includegraphics[width=\hsize]{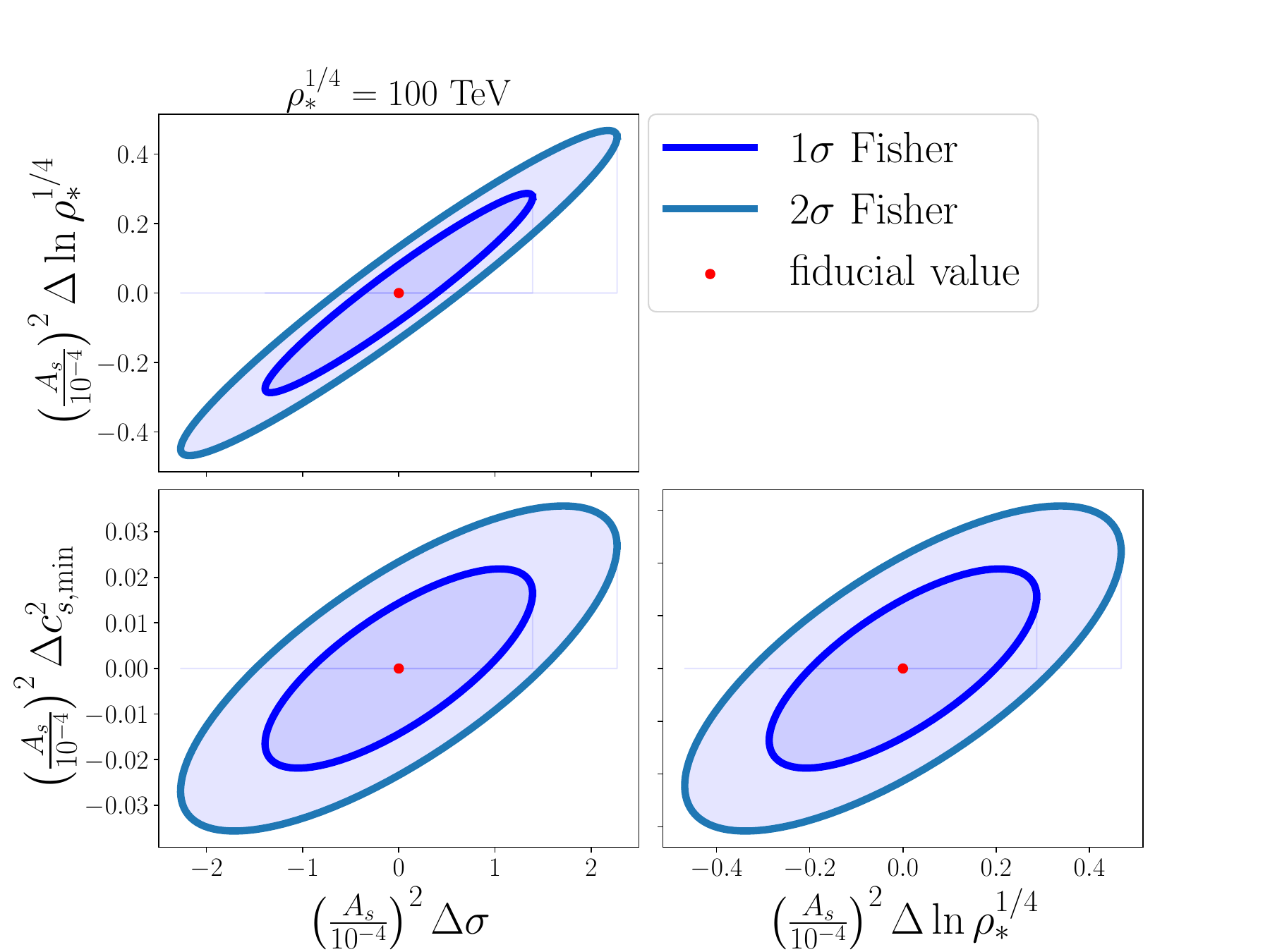}
        \end{minipage}
        \begin{minipage}{0.5\hsize}
            \centering
            \includegraphics[width=\hsize]{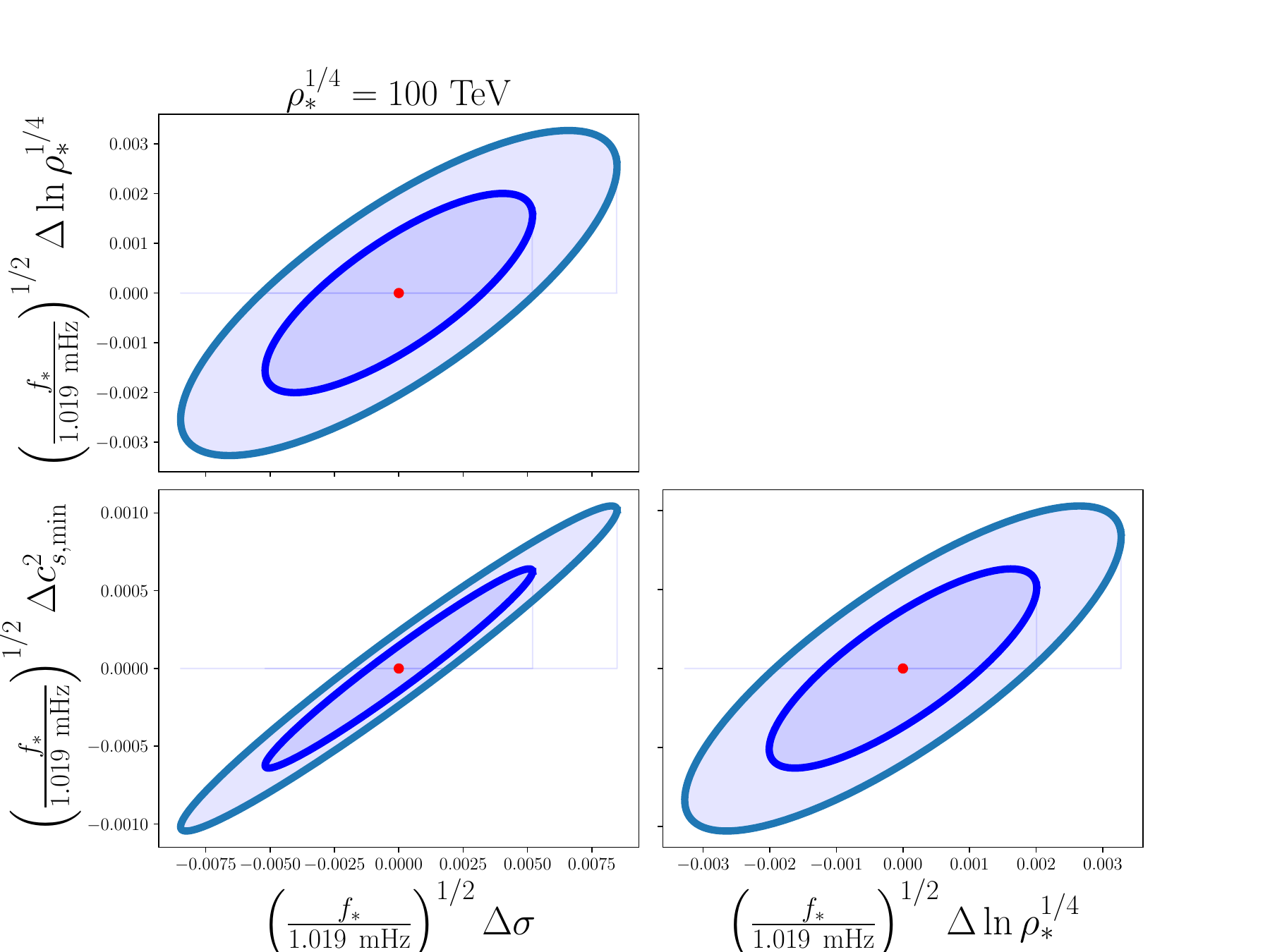}
        \end{minipage}
    \end{tabular}
    \caption{$1\sigma$ (bright blue) and $2\sigma$ (pale blue) confidence ranges for the parameter variations $\Delta\ln\rho_*^{1/4}$, $\Delta\csmin^2$, and $\Delta\sigma$ round the fiducial value $(\rho_*^{1/4},\csmin^2,\sigma)=(\SI{100}{TeV},0.1,2)$ in the weak limit~\eqref{eq: F weak} (left) and the strong limit~\eqref{eq: F strong} (right).}
    \label{fig: Fisher weak and strong}
\end{figure*}

With the use of the \ac{GW} spectral derivatives in the model parameters, the Fisher information matrix for \acp{SGWB} is calculated as~\cite{Tegmark:1996bz, Smith:2019wny, Boileau:2021gbr} 
\bae{\label{eq: fisher_full_component}
    F_{ij} 
    &= \frac{1}{2}T_\obs \sum_{I}\int^{f_h}_{f_l} \dd{f} 
    \frac{\partial_{\theta_i}
    \qty(\Omega_\GW(f)h^2)
    \partial_{\theta_j}
    \qty(\Omega_\GW(f)h^2)}{\qty[\Omega_{n,II}(f)h^2 + \Omega_\GW(f)h^2]^2},
}
where $T_\obs$ denotes the observational time and $\Omega_{n,II}$ is the noise density spectrum in the independent $I$ channel of the \ac{TDI} measurement.
Explicit expressions for $\Omega_{n,AA}h^2=\Omega_{n,EE}h^2$ and $\Omega_{n,TT}h^2$ for three independent $A$, $E$, and $T$ channels of LISA are given in the Appendix~\ref{sec: LISA noise}. 
See Fig.~\ref{fig: Omg_vs_LISA} for explicit shapes of the noise density spectra of \ac{LISA}.
The integration range of the Fisher matrix is set to $f_l=10^{-2}f_*$ and $f_h=10^2f_*$. The covariance matrix is given by the inverse of the Fisher matrix (see, e.g., Ref.~\cite{Coe:2009xf}).

\begin{figure*}
    \centering
    \includegraphics[width=0.8\hsize]{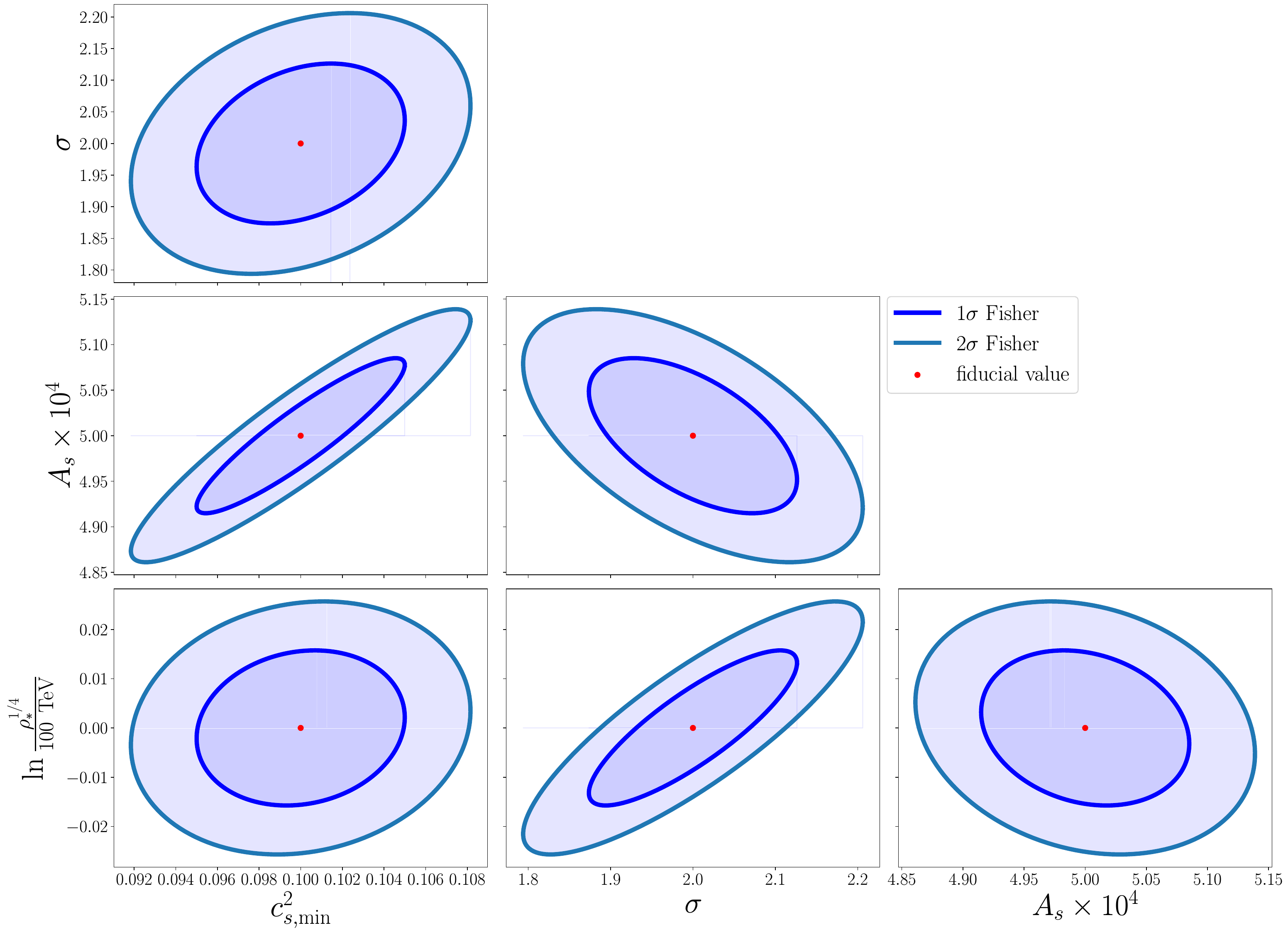}
    \caption{The marginalized Fisher forecast with the exact expression~\eqref{eq: fisher_full_component} assuming the one-year observation. The red point denotes the fiducial values, $(\csmin^2,\,\sigma,\,A_\us,\, \rho^{1/4}_*) = (0.1,\,2.0,\,5 \times 10^{-4},\,\SI{100}{TeV})$. 
    }
    \label{fig: fisher_SIGW_marginalized}
\end{figure*}

In the weak-signal limit, $\Omega_\GW h^2\ll\Omega_{n,II}h^2$, the parameter dependence of the Fisher matrix can be factorized as
\bme{\label{eq: F weak}
    F_{ij}\underset{\text{weak}}{\sim}\num{2.84e21}\pqty{\frac{T_\obs}{\SI{1}{yr}}}\pqty{\frac{\Omega_{\ur0}h^2}{\num{4.2e-5}}}^2\pqty{\frac{A_\us}{10^{-4}}}^4 \\
    \times\pqty{\frac{f_*}{\SI{1.019}{mHz}}}\sum_I\int_{\tilde{f}_l}^{\tilde{f}_h}\dd{\tilde{f}}\frac{\partial_{\theta_i}\calT_\GW(\tilde{f})\partial_{\theta_j}\calT_\GW(\tilde{f})}{\pqty{\Omega_{n,II}(\tilde{f}f_*)h^2}^2},
}
where $\tilde{f}_l=10^{-2}$ and $\tilde{f}_h=10^2$.
The parameter uncertainties hence scale as $\propto A_\us^{-2}$ as expected.\footnote{Note that the $f_*$ dependence is not simply linear as the noise spectrum depends on $f_*$ in the integral.}
The confidence ellipses are shown in the left panel of Fig.~\ref{fig: Fisher weak and strong} around $(\rho_*^{1/4},\csmin^2,\sigma)=(\SI{100}{TeV},0.1,2)$ in this limit.
While the \ac{SIGW} is not so sensitive to the crossover smoothness $\sigma$ in the weak-signal limit, its scale $\rho_*$ and strongness
$\csmin^2$ can be determined at $\sim10\%$ level for $A_\us\sim10^{-4}$ which is smaller enough than the amplitude required for a sizable \ac{PBH} formation, $A_\us\sim\num{2e-3}$~\cite{Escriva:2023nzn}.

In the strong-signal limit, $\Omega_\GW h^2\gg\Omega_{n,II}h^2$, the Fisher matrix becomes free from the scalar amplitude $A_\us$ as well as the detector design and is determined only by the \ac{GW} template form and the characteristic frequency $f_*$ normalized by the observation time $T_\obs$:
\bme{\label{eq: F strong}
    F_{ij}\underset{\text{strong}}{\sim}\num{4.82e4}\pqty{\frac{T_\obs}{\SI{1}{yr}}}\pqty{\frac{f_*}{\SI{1.019}{mHz}}} \\
    \times\int_{\tilde{f}_l}^{\tilde{f}_h}\dd{\tilde{f}}\frac{\partial_{\theta_i}\calT_\GW(\tilde{f})\partial_{\theta_j}\calT_\GW(\tilde{f})}{\calT_\GW^2(\tilde{f})}.
}
The corresponding confidence ellipses are shown in the right panel of Fig.~\ref{fig: Fisher weak and strong}. It shows the strongest possible decidability of the crossover parameters with the use of \acp{SIGW} irrespective of the \ac{GW} detector.
One can see the significant degeneracy between $\csmin^2$ and $\sigma$, which makes the decidability of each of $\csmin^2$ and $\sigma$ worse.
As the denominator in the integrand of the Fisher matrix is dominated by the \ac{GW} spectrum in this limit, the Fisher matrix is sensitive not only to the large-($\partial_{\theta_i}\calT_\GW$) part but also to the small-$\calT_\GW$ part.
Thus the frequency range much higher than $f_*$ also significantly contributes to the Fisher matrix. 
Since both the strongness $\csmin^2$ and the width $\sigma$ contribute to the reduction of $\calT_\GW$ in the higher frequency range, we find the significant degeneracy in the strong-signal limit. 
That is, a less significant reduction $\Delta\csmin^2>0$ can be compensated by a longer crossover $\Delta\sigma>0$ and vice versa.

In Fig.~\ref{fig: fisher_SIGW_marginalized}, the full confidence ellipses including $A_\us$ (note $\partial_{A_\us}\Omega_\GW h^2=2\Omega_\GW h^2/A_\us$) are calculated from the exact expression~\eqref{eq: fisher_full_component} as our main result.

\section{Conclusions}

We exhibited the Fisher forecast on the model parameters of the phenomenological sound-speed reduction~\eqref{eq: cs2} in the early universe associated with a certain smooth crossover around $\SI{100}{TeV}$ beyond the Standard Model of particle physics through the \ac{SIGW} supposing the one-year \ac{LISA} observation.
The parameter decidability depends on the amplitude $A_\us$ of the primordial scalar perturbations around the corresponding scale $\sim10^{12}\,\si{Mpc^{-1}}$ as a source of \acp{SIGW}.
The dependence can be inferred by the left (weak $A_\us$ limit) and the right (strong $A_\us$ limit) of Fig.~\ref{fig: Fisher weak and strong}.
The main forecast is shown in Fig.~\ref{fig: fisher_SIGW_marginalized}, indicating that $A_\us\sim\num{5e-4}$ is enough to determine all model parameters.

If the scalar amplitude is slightly higher as $A_\us\sim10^{-3}$, sizble amount of \acp{PBH} can be produced at $\sim10^{22}\,\si{g}$ lying at the observational window for whole dark matters~\cite{Escriva:2022yaf,Escriva:2023nzn}.
A significant enough crossover makes the \ac{PBH} mass function sharp enough to avoid all observational constraints on \acp{PBH} even if the scalar power spectrum is almost scale-invariant as long as the amplitude is not large enough for a sizable amount of \acp{PBH} without a crossover (see also, e.g., Refs.~\cite{Byrnes:2018clq,Carr:2019kxo} for a similar phenomenon with the Standard Model crossovers).
Combining our work with this scenario, one can determine not only the crossover parameters but also the abundance of \acp{PBH}, as indirect evidence of the nature of dark matter if they are (of course under the assumption of the Gaussianity of the primordial perturbation though).

The Fisher analysis relies on the assumption of the Gaussianity of the \ac{GW} signal, while one may be concerned that \acp{SIGW} are intrinsically non-Gaussian.
Fortunately, it has been proven in Ref.~\cite{Bartolo:2018evs} that the propagation effects on the \ac{SGWB} make it completely Gaussian.
Our analysis is also based on several further assumptions such as the scale invariance~\eqref{eq: calPzeta} and the Gaussianity of the scalar perturbations and also no other comparable \ac{GW} source represented by the coalescence of supermassive black holes.
Though our qualitative result is expected not to be significantly altered unless some of them are obviously violated, it is definitely of physical interest to relax these assumptions in the Fisher forecast (see, e.g., Ref.~\cite{Babak:2024yhu} for a generalization to a non-flat spectrum and a subdominant \acp{SIGW} for \ac{PTA} experiments).  
We leave it for future work.

\acknowledgments

We are grateful to Sachiko Kuroyanagi and Javier G. Subils for their helpful discussions. 
Y.T. is supported by JSPS KAKENHI Grants
No.~JP21K13918 and JP24K07047.
R.I. is supported by JST SPRING, Grant Number JPMJSP2125, and the ``Interdisciplinary Frontier Next-Generation Researcher Program of the Tokai Higher Education and Research System". 
A.E. acknowledges support from the JSPS Postdoctoral Fellowships for Research in Japan (Graduate School of Sciences, Nagoya University).
C.Y. is supported by JSPS KAKENHI Grants No.~JP20H05850 and JP20H05853.

\appendix
\begin{widetext}

\section{\boldmath Crossover scale $\rho_*$ and the characteristic frequency $f_*$}\label{sec: rho and f}

The characteristic frequency $f_*=k_*/(2\pi)=\calH_*/(2\pi)$, where $\calH_*=a_*H_*$ is the conformal Hubble parameter at the time when $\rho=\rho_*$, can be found by numerically solving the background equation (Eq.~(4) in the main document) with the normalization $a=1$ today.
Supposing the standard cosmology in the Standard Model of particle physics is recovered at, e.g., $\rho^{1/4}=10\,\si{TeV}$, The normalization can be recast into the condition $a=\num{1.93e-17}$ and $\eta=\SI{1.40e-11}{Mpc}$ at $\rho^{1/4}=10\,\si{TeV}$ with the use of the fitting formulae for effective degrees of freedom given in Ref.~\cite{Saikawa:2018rcs}.
For the fiducial values $(\rho_*^{1/4},\csmin^2,\sigma)=(\SI{100}{TeV},0.1,2)$, one finds $f_*=\SI{1.019}{mHz}$.

The $f_*$ dependence on these parameters around the fiducial values can be numerically found by slightly variating the parameters.
Fig.~\ref{fig: fs dependence} shows numerically calculated $f_*$ with shifted parameters as well as their fitting functions.
Combining them, one obtains the fitting formula (Eq.~(12) in the main document).

\begin{figure}
    \centering
    \begin{tabular}{c}
        \begin{minipage}{0.33\hsize}
            \centering
            \includegraphics[width=\hsize]{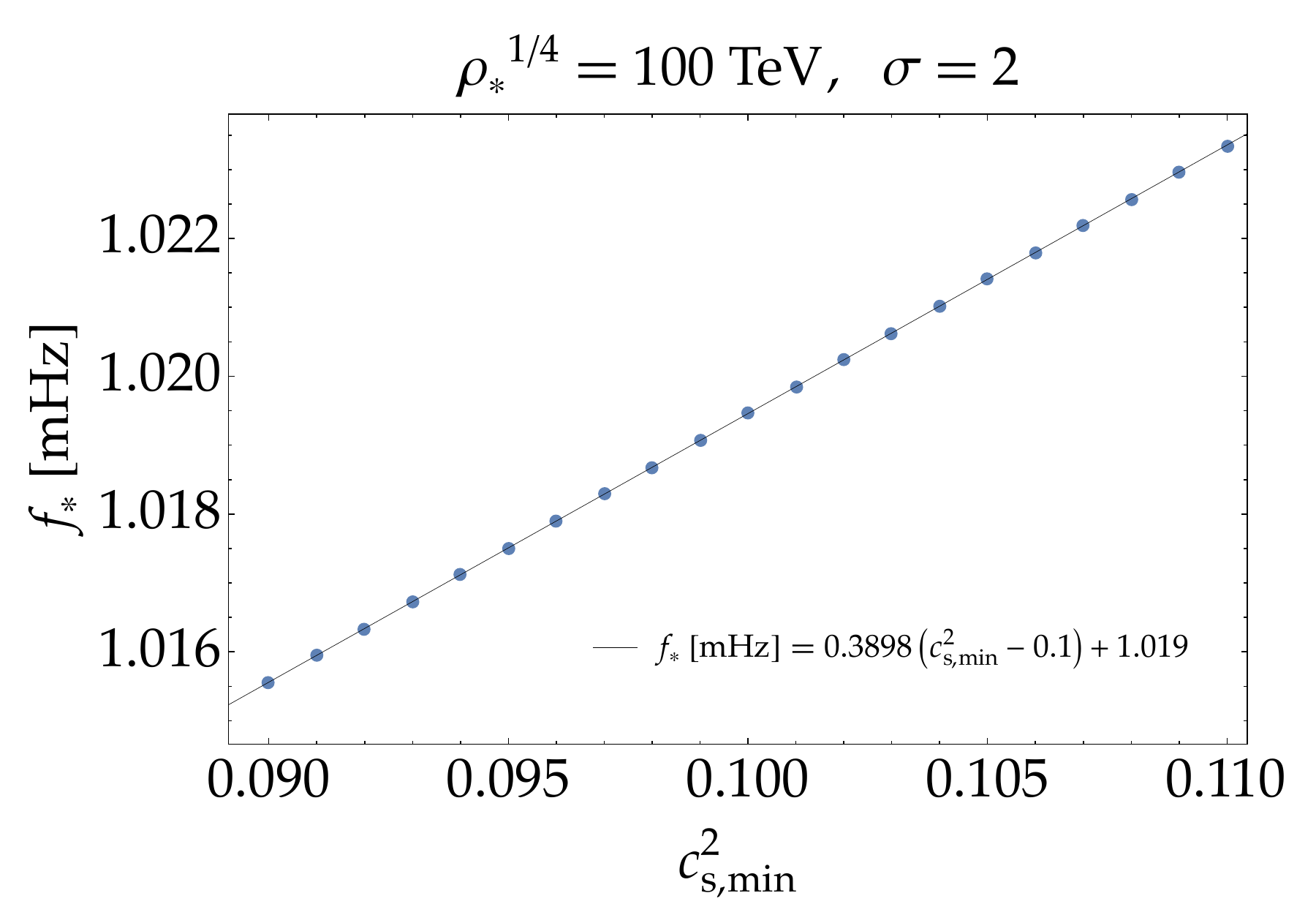}
        \end{minipage}
        \begin{minipage}{0.33\hsize}
            \centering
            \includegraphics[width=\hsize]{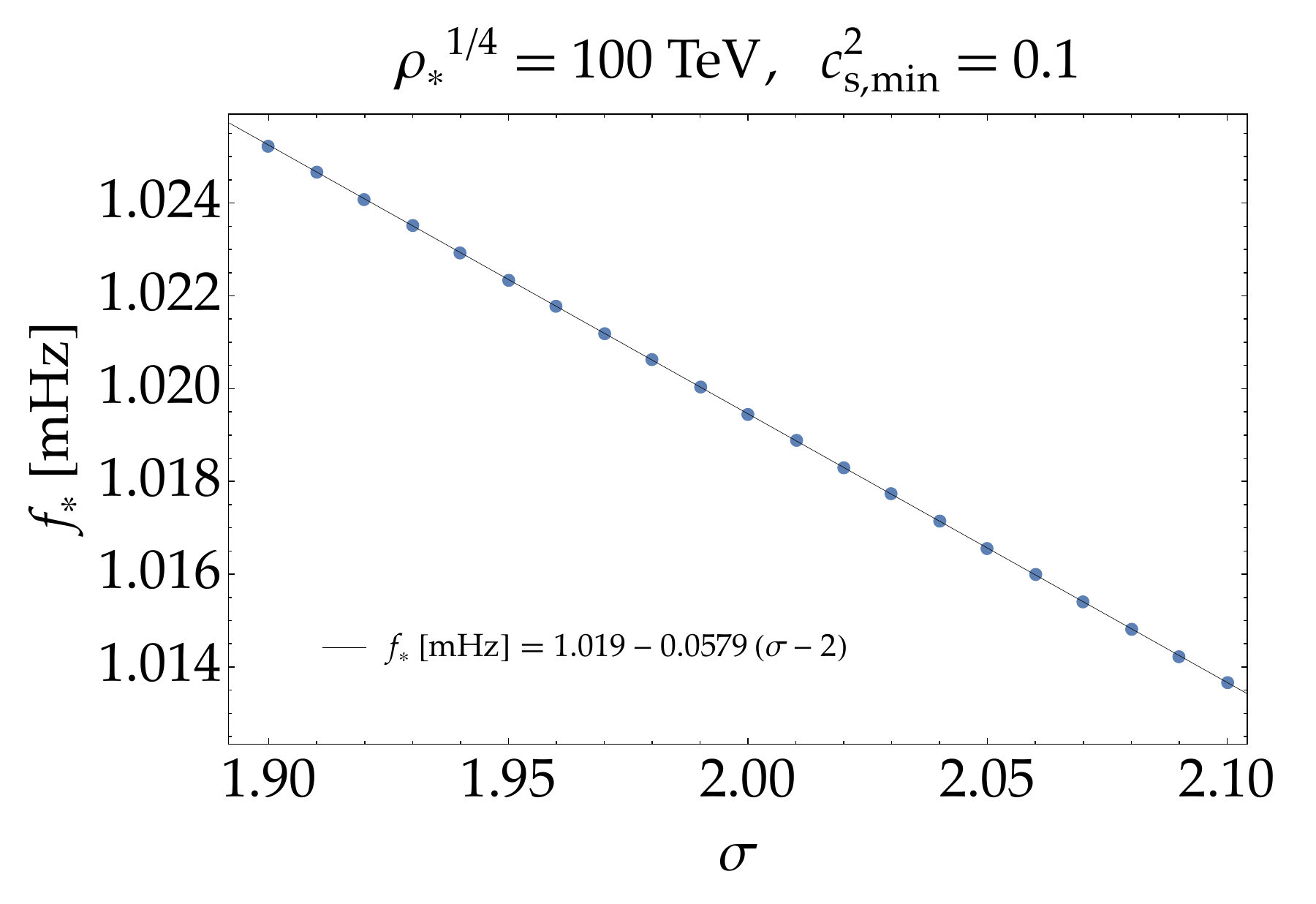}
        \end{minipage}
        \begin{minipage}{0.33\hsize}
            \centering
            \includegraphics[width=\hsize]{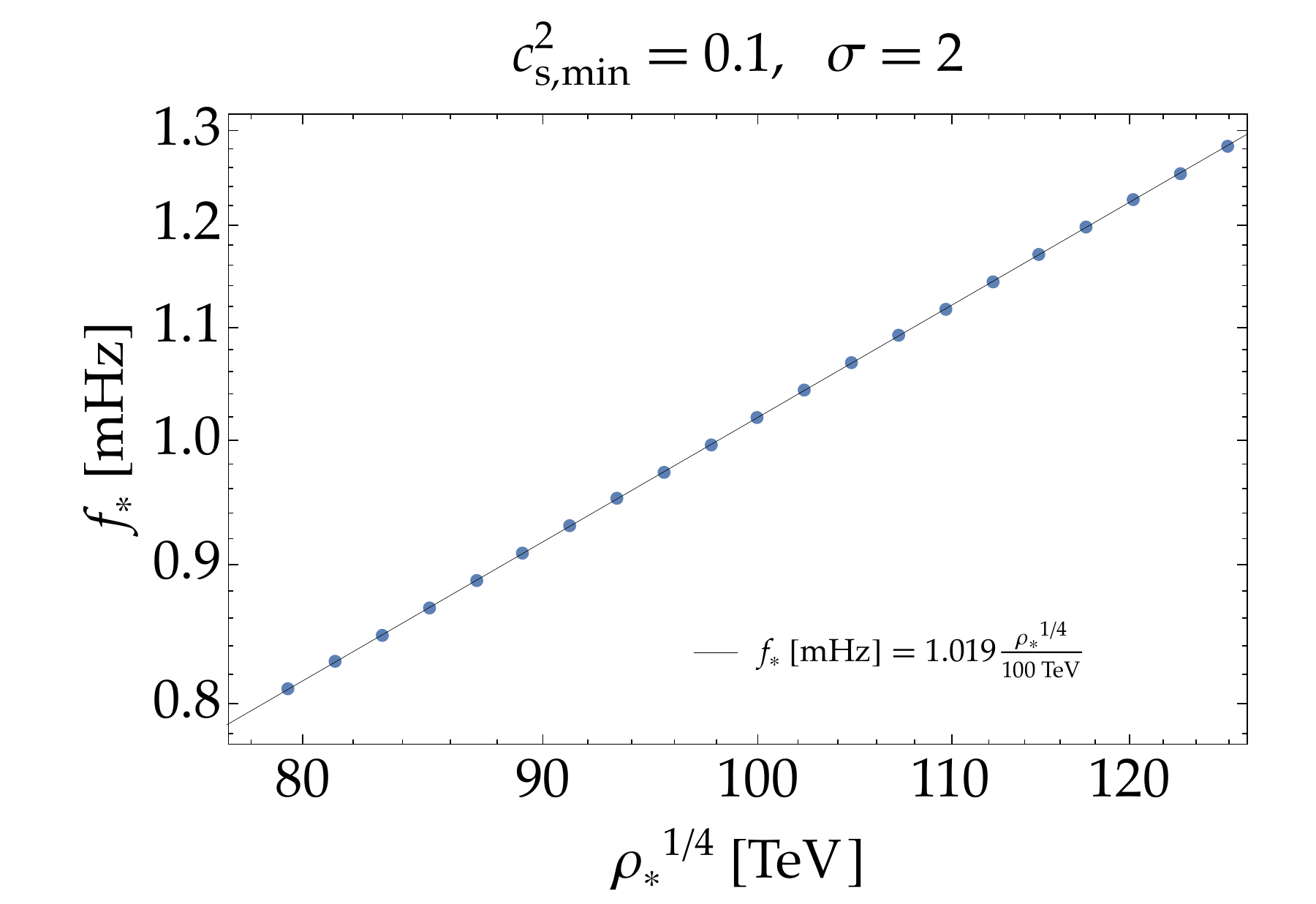}
        \end{minipage}
    \end{tabular}
    \caption{The $f_*$ dependence on $\csmin^2$ (left), $\sigma$ (middle), and $\rho_*$ (right). Blue dots are the numerical results and thin lines are their fitting functions. The fitting formula (Eq.~(12) in the main document) combines them all.}
    \label{fig: fs dependence}
\end{figure}

\section{Noise density parameters of LISA}\label{sec: LISA noise}

The noise density parameters of LISA are summarized as~\cite{Smith:2019wny, Flauger:2020qyi}
\bae{
    \Omega_{n,II}(f)h^2=\frac{4\pi^2f^3}{3\times(\SI{100}{km/s/Mpc})^2}\frac{N_{II}(f)}{\calR_{II}(f)},
}
where 
\beae{
    N_{AA}(f) &= N_{EE}(f) \\
    &= 8 \sin[2](\frac{2\pi f L}{c})\Bqty{4\qty[1 + \cos(\frac{2\pi f L}{c}) + \cos[2](\frac{2\pi f L}{c})]P_{\mathrm{acc}}(f) + \qty[2 + \cos(\frac{2\pi f L}{c})]P_{\mathrm{IMS}}(f) }, \\
    N_{TT}(f) &= 16 \sin[2](\frac{2\pi f L}{c})\Bqty{
    2\qty[1 - \cos(\frac{2\pi f L}{c})]^2P_{\mathrm{acc}}(f) + \qty[1 - \cos(\frac{2\pi f L}{c})]P_{\mathrm{IMS}}(f)},
}
and
\bae{
    \calR_{IJ}(f) = 16\sin[2](\frac{2 \pi f L}{c}) \qty(\frac{2 \pi f L}{c})^2 \tilde{R}_{IJ}(f),
}
with 
\beae{\label{eq: noise power}
    &P_{\mathrm{IMS}} =\qty(P\,\si{pm.Hz^{-1/2}})^2\qty[1 + \qty(\frac{\SI{2}{mHz}}{f})^4]\qty(\frac{2\pi f}{c})^2, \\ 
    &P_{\rm acc} = \qty(A\,\si{fm.s^{-2}.Hz^{-1/2}})^2\qty[1 + \qty(\frac{\SI{0.4}{mHz}}{f})^2]\qty[1 + \qty(\frac{f}{\SI{8}{mHz}})^4]\qty(\frac{1}{2\pi f})^4\qty(\frac{2\pi f}{c})^2, 
}
and
\bae{
    \tilde{R}_{AA}(f) = \tilde{R}_{EE}(f) =  \frac{9}{20} \frac{1}{1 + 0.7\left(\frac{2 \pi f L}{c}\right)^2 } \qc
    \tilde{R}_{TT}(f) = \frac{9}{20} \frac{\left(\frac{2 \pi f L}{c}\right)^6}{1.8 \times 10^3 + 0.7\left(\frac{2 \pi f L}{c}\right)^8 }.
}
We explicitly show the speed of light $c$ for clarity.
The dominant noise components of the \ac{LISA} system are the position noise so-called the \ac{IMS} due to laser shot noise and the acceleration noise raised from internal sensors.
$P=15$ and $A=3$ are chosen for the noise power~\eqref{eq: noise power}, following Refs.~\cite{Smith:2019wny, Flauger:2020qyi}.

\end{widetext}

\bibliography{main}

\end{document}